\documentclass[tradiabstract]{aa} % for the abstract without structuration 

\usepackage{graphicx}
\usepackage{txfonts}
\usepackage{natbib}
\bibpunct{(}{)}{;}{a}{}{,}
\begin{document}
   \title{HAWK-I imaging of the X-ray luminous galaxy cluster
   \object{XMMU\,J2235.3-2557}\thanks{Based on observations obtained
   at the European Southern Observatory using the ESO Very Large
   Telescope on Cerro Paranal through ESO program 060.A-9284(H).}}

   \subtitle{The red sequence at $z=1.39$}

   \author{C. Lidman\inst{1}
          \and
	  P. Rosati\inst{2}
          \and
	  M. Tanaka\inst{2}
	  \and
	  V. Strazzullo\inst{3}
	  \and
	  R. Demarco\inst{4}
	  \and
	  C. Mullis\inst{5}
	  \and
	  N. Ageorges\inst{6}
	  \and
	  M. Kissler-Patig\inst{2}
	  \and
	  M. G. Petr-Gotzens\inst{2}
	  \and
	  F. Selman\inst{1}
          }

   \institute{European Southern Observatory, Alonso de Cordova 3107,
              Casilla 19001, Santiago, Chile
         \and
             European Southern Observatory, Karl-schwarzschild-Strasse 2,
              D-85748 Garching, Germany
	  \and
	     National Radio Astronomy Observatory, P.O. Box O, Socorro, NM, 87801, USA
	  \and
             Department of Physics and Astronomy, John Hopkins University, Baltimore, USA
	  \and
	     Wachovia Corporation, NC6740, 100 N. Main Street, Winston-Salem, NC 27101
	  \and
	     MPI f\"ur extraterrestische Physik, Giessenbachsrtasse, D-85748 Garching, Germany
          }

   \date{Received July 9, 2008; accepted July 31, 2008}
                                   
% \abstract{}{}{}{}{} 
% 5 {} token are mandatory

  \abstract {We use HAWK-I, the recently-commissioned near-IR imager
    on Yepun (VLT-UT4), to obtain wide-field, high-resolution images
    of the X-ray luminous galaxy cluster \object{XMMU\,J2235.3-2557}
    in the J and Ks bands, and we use these images to build a
    colour-magnitude diagram of cluster galaxies. Galaxies in the core
    of the cluster form a tight red sequence with a mean
    $\mathrm{J-Ks}$ colour of 1.9 (Vega system). The intrinsic scatter
    in the colour of galaxies that lie on the red sequence is similar
    to that measured for galaxies on the red sequence of the Coma
    cluster.  The slope and location of the red sequence can be
    modelled by passively evolving the red sequence of the Coma
    cluster backwards in time. Using simple stellar population (SSP)
    models, we find that galaxies in the core of
    \object{XMMU\,J2235.3-2557} are, even at $z=1.39$, already 3\,Gyr
    old, corresponding to a formation redshift of $z_{\rm f} \sim 4$. Outside
    the core, the intrinsic scatter and the fraction of galaxies
    actively forming stars increase substantially. Using SSP models,
    we find that most of these galaxies will join the red sequence
    within 1.5\,Gyr. The contrast between galaxies in the cluster core
    and galaxies in the cluster outskirts indicates that the red
    sequence of \object{XMMU\,J2235.3-2557} is being built from the
    dense cluster core outwards.}


   \titlerunning{The red sequence of XMMU\,J2235.3-2557}
   \authorrunning{C. Lidman et al.}

   \maketitle
%
%________________________________________________________________

\section{Introduction}

The most notable feature in the colour-magnitude (C-M) diagram of a
rich galaxy cluster is the tight red sequence of passively-evolving
galaxies. First observed in nearby galaxy clusters
\citep[e.g.,][]{Vaucouleurs61,Visvanathan77}, it has also been
observed in some of the most distant galaxy clusters currently known
\citep[e.g.,][]{vanDokkum01,Blakeslee03,Lidman04,Holden04,Mei06}.

The location and tightness of the red sequence are used to infer
that the bulk of the stars in red sequence galaxies formed over a
relatively short space of time at much higher redshifts
\citep{Bower92,Stanford98}. In \object{RDCS~J1252-2927} at $z=1.24$,
for example, \citet{Blakeslee03} find that most of the stars in
the galaxies that lie on the red sequence formed 11.9 Gyr ago.

The red sequence is tilted in the sense that brighter galaxies are
also redder. The tilt is understood as a relationship between mass and
metallicity \citep{Kodama99,Gallazzi06}. Brighter, more massive
galaxies are more metal rich and are hence redder. The tilt is
observed at all redshifts, and there is little evidence that it
evolves by more than that expected from passive evolution of an old
stellar population \citep{Stanford98,Blakeslee03}. However, there are
hints that it flattens at the bright end for some $z > 1$ clusters
\citep{vanDokkum01,Demarco07} and for S0 galaxies in the cluster
\object{RDCS~J0910+5422} at $z=1.106$ \citep{Mei06}.

When sufficiently precise measurements have been made, the sequence
has a small, but non-zero, scatter in colour
\citep{Bower92,Eisenhardt07}.  For the most massive clusters, the
scatter appears to be independent of redshift
\citep{Stanford98,vanDokkum01,Blakeslee03,Mei06,Homeier06}; however,
in some less massive clusters \citep{Holden04}, the scatter is larger
and the red sequence is less regular. While it is generally accepted
that the tilt in the colour-magnitude relation is primarily the result
of a relationship between mass and metallicity, the reasons for the
scatter are less clear. For early-type galaxies in the field, the
scatter is caused by differences in age and metallicity, with
variations in metallicity playing an increasingly important role as
the mass of the galaxy increases \citep{Gallazzi06}. Dust could also
play a role.

Within the framework of hierarchical structure formation, one expects
the morphology of the red sequence to evolve as one approaches the
epoch of cluster formation. The major processes are thought to be:
passive evolution of old stellar populations, conversion of galaxies
that are in the ``blue cloud'' into red ones through the quenching of
star formation, and dry mergers \citep{Bell06,Faber07}. The time at
which these processes initiate and the speed at which they run depend
on environment, with the result that the red sequences of the most massive
clusters form first \citep{Tanaka05,Romeo08,Menci08}.

Recent observations of the most distant clusters have started to note
changes in the morphology of the red sequence. A deficit of galaxies
at the faint end of the red sequence has been noted in the large scale
structure surrounding the high redshift cluster
\object{RDCS~J1252-2927} at $z=1.24$ \citep{Tanaka07}. A similar
deficit in the cluster itself was not seen, suggesting that the rate
at which the red sequence develops depends on environment, as one
expects in hierarchical models and as seen in some clusters at lower
redshifts \citep{Tanaka05}. The faint end deficit has been noted in
clusters from the ESO distant cluster survey
\citep{deLucia04b,deLucia07}; however, no such deficit was found in
the cluster sample that was analysed by \citet{Andreon08}, perhaps
indicating that the clusters analysed by Andreon were, on average,
richer than those in ESO distant cluster survey. A similar richness
dependence was noted in clusters from the Red Cluster Survey
\citep{Gilbank08}.

In a recent study of several proto-clusters with redshifts ranging
from $z=2.2$ to $z=3.1$, \citet{Kodama07} found that the bright end of
the red sequence becomes progressively less well defined as the
redshift of the proto-cluster increases. In one of the most
well-studied proto-clusters, \object{PKS~1138-262} at $z=2.16$,
\citet{Zirm08} found that the red sequence of \object{PKS~1138-262} is
considerably less well defined than the red sequence of rich clusters at
lower redshifts.

Hence, between the redshift of most well studied clusters at $z \sim
1.3$ and the most well studied proto-clusters at $z \sim 2$, there is
evidence for a dramatic change in the morphology of the red sequence.
Massive X-ray luminous clusters beyond a redshift of 1.3 are rare,
and currently, only 2 are known, \object{XMMU~J2235.3-2557}
\citep{Mullis05} at $z=1.39$ and \object{XMMXCS~J2215.9-1738} at
$z=1.45$ \citep{Stanford06}.

In this paper, we present the near-IR colour-magnitude (C-M) diagram
of \object{XMMU~J2235.3-2557}, one of the most distant X-ray luminous
clusters currently known. In section 2 of the paper, we describe the
near-IR observations and the methods used to process the data. In
section 3, we present the C-M diagram and derive the intrinsic
scatter in the colour of galaxies that lie on the red sequence. In
section 4, we discuss the results. Throughout this paper, we assume a
flat, $\Lambda$-dominated universe with $\Omega_{\mathrm M}=0.27$ and
$H_{\mathrm 0} = 71\,\mathrm{km\,s^{-1}\,Mpc^{-1}}$. In this
cosmology, 1\arcsec\ on the sky corresponds to 8.5\,kpc at
$z=1.39$. Unless specified otherwise, all magnitudes are Vega
magnitudes and are on the 2MASS system. In a forthcoming paper, we
will present an analysis of multi-wavelength observations of
\object{XMMU~J2235.3-2557}, including spectroscopy from VLT/FORS2 and
optical imaging from ACS/HST.

%__________________________________________________________________

\section{Observations and Photometry}

\subsection{Observations}

\object{XMMU~J2235.3-2557} was observed with HAWK-I
\citep[][Kissler-Patig et al. in preparation]{Pirard04,Casali06} on
Yepun (VLT-UT4) at the ESO Cerro Paranal Observatory. The observations
were taken in service mode during the first two weeks of October 2007
as part of the first HAWK-I science verification run.

HAWK-I is a near-IR imager with a 7\farcm5 x 7\farcm5 field of
view. The focal plane consists of a mosaic of 4 Hawaii-2RG detectors
and results in an average pixel scale of 0\farcs1065 per pixel. The
cluster was imaged in J and Ks.

In order to avoid the gaps between the detectors in the mosaic and to
cover a wide area, the observations were not done with the cluster
positioned in the centre of the mosaic. Instead, a series of four
pointings with the cluster positioned in the four outer corners of the
mosaic was used. The resulting union of images covers 13\farcm5 by
13\farcm5 of the sky without gaps.

Individual exposures lasted 10 seconds in both J and Ks,
and 12 (6 for Ks) of these were averaged to form a single
image. Between images, the telescope was moved by 10\arcsec\ to
30\arcsec\ in a semi-random manner, and 22 to 45 images were taken in
this way in a single observing block. In the cluster centre, which was
always imaged, the total exposure times for J and Ks were 176 and 179
minutes, respectively. Exposure times and detection limits are
reported in Table \ref{table:observations}.

Standards were selected from the LCO \citep{Persson98} and UKIRT
\citep{Hawarden01} standard star lists. The UKIRT standards were used
to monitor the transparency of the nights, which were stable to within
2\% during the period the cluster was observed. The LCO standards were
used to set the zero point.

\subsection{Data Processing\label{sec:data}}

The processing of the raw data was done in a standard manner and
consisted of the following steps:

\begin{enumerate}

\item subtraction of dark frames to remove the zero-level offset

\item division by normalised twilight flats to normalise the
pixel-to-pixel response

\item object masked sky subtraction using the XDIMSUM package in
  IRAF\,\footnote{IRAF is distributed by the National Optical Astronomy
  Observatories which are operated by the Association of Universities
  for Research in Astronomy, Inc., under the cooperative agreement
  with the National Science Foundation}

\item normalisation of the detector gains using the instrumental
magnitudes of the UKIRT and LCO standards.

\item astrometric calibration using SExtractor (version 2.5.0) and
  SCAMP (version 1.4.0)\,\footnote{http://terapix.iap.fr}

\item image combination using SWarp (version 2.16.4)\,$^2$

\end{enumerate}

The mean count level in the raw images varies from 2,000 ADU in 10
seconds for J to 10,000 ADU in 10 seconds for Ks. Since the 1\%
non-linearity threshold is 30,000 ADU, corrections for detector
non-linearity were not applied.

The accuracy of the flat fielding over the HAWK-I field of view was
tested by observing stars over a 5x5 or 9x9 grid.  For sufficiently
bright stars, the dispersion in the instrumental magnitude of any one
star was never greater than 0.02 magnitudes.

The accuracy of the zero points were assessed by comparing the
magnitudes of stars in the HAWK-I images with the magnitudes listed in
the 2MASS point source catalogue \citep{Skrutskie06}. For stars in the
magnitude range $13.5 < J_\mathrm{2MASS} < 15.5$ ($13 <
Ks_\mathrm{\,2MASS} < 15$ for Ks) the difference
$J_\mathrm{HAWK-I}-J_\mathrm{2MASS}$, averaged over 26 stars, was
0.025 magnitudes (for Ks, $Ks_\mathrm{HAWK-I}-Ks_\mathrm{2MASS}=0.010$
magnitudes, averaged over 30 stars). Stars brighter than 13.5\,magnitudes in
J and 13\,magnitudes in Ks were saturated in the HAWK-I images.

The fully processed frames were co-added to produce two sets of
images. In the first set, all frames with a given filter were co-added
to produce a deep image of the centre of the cluster. In the second
set, the frames corresponding to a single pointing were co-added
together. With four pointings, this results in four images per filter.
In both sets and for the purpose of maximising image quality,
individual frames are weighted by the square of the inverse of the
FWHM of stars in those frames. This results in five pairs of images,
one for the centre of the cluster and four for the cluster
outskirts. Each image pair is then processed (the image with the best
image quality is smoothed with a Moffat function) so that the image
quality within that pair match. The separation between the core of the
cluster, which is common to all frames, and the cluster outskirts,
which are not common to all frames, facilitates the task of computing
colours from aperture magnitudes.

\begin{table}
\caption{Exposure times, image quality and detection limits of the
reduced HAWK-I data.} 
\label{table:observations} 
\centering
\begin{tabular}{c c c c} 
\hline\hline
Filter & Exposure time$^{\mathrm{a}}$ & Image quality & Detection limit$^{\mathrm{b}}$ \\
       & (seconds)     & (\arcsec)     & (Vega magnitudes)\\
\hline
J      & 10560         & 0.47          & 24.9 \\
Ks     & 10740         & 0.32          & 23.2 \\
\hline
\end{tabular}

\begin{list}{}{}
 
\item[$^{\mathrm{a}}$] All quantities refer to the central part of the
mosaic, where exposure times are greatest.

\item[$^{\mathrm{b}}$] The detection limit is the 5
sigma point source detection limit within an aperture that has a
diameter equal to twice the image quality.

\end{list}

\end{table}

\begin{figure*}
\centering
\includegraphics[width=17cm]{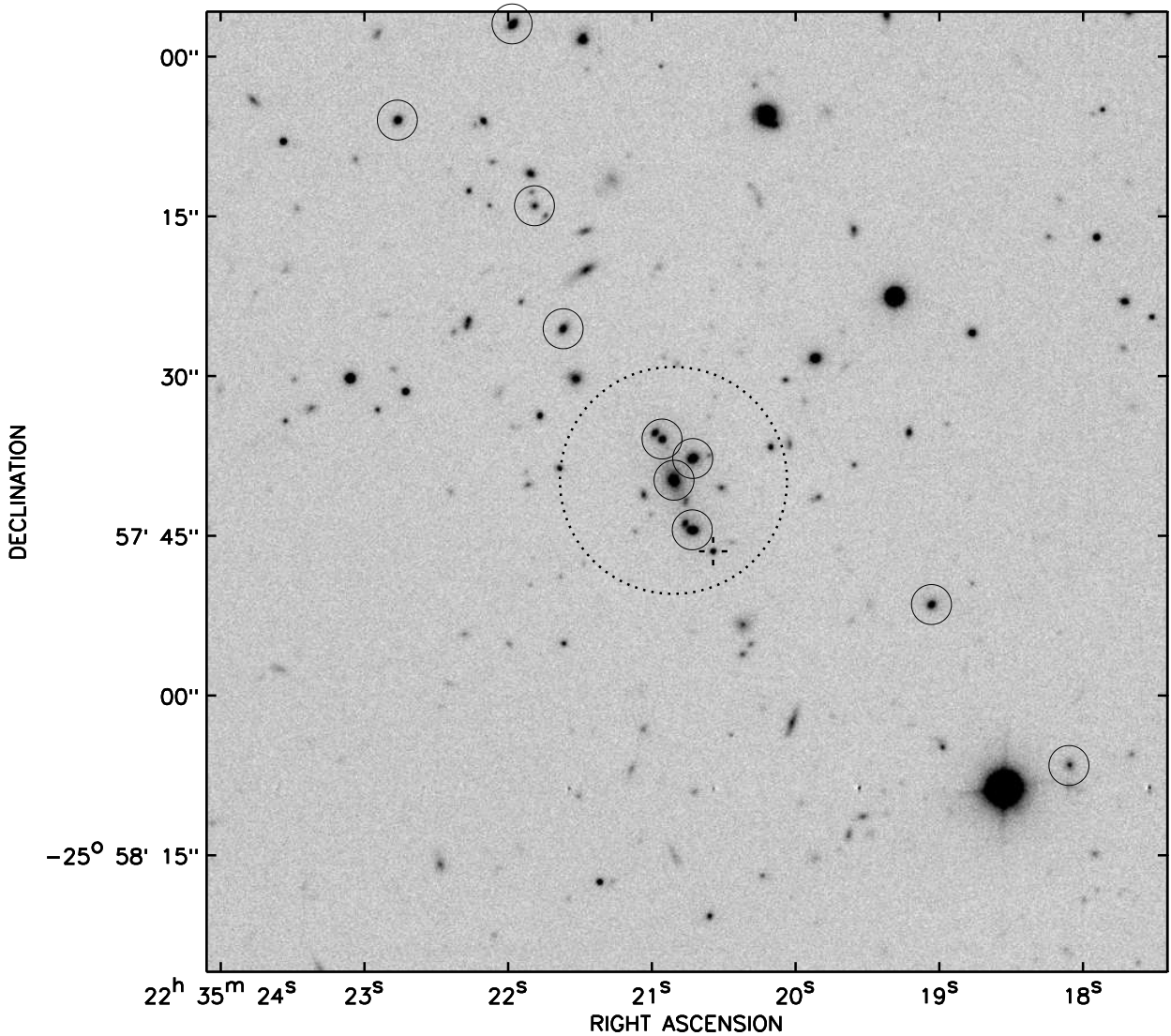}

\caption{HAWK-I Ks-band image of \object{XMMU\,J2235.3-2557}. Only the
  central 1\% of the entire Ks-band image is shown here. The image is
  approximately 90\,\arcsec\ on a side, which, in the adopted
  cosmology, corresponds to 765 kpc at $z=1.39$. Spectrally confirmed
  cluster members (Rosati et al. in preparation) are marked with
  circles. The star that is close to the cluster centre is also
  marked. The dotted circle that surrounds the central region of the
  cluster has a diameter of 21\,\arcsec, which corresponds to 180\,kpc
  at $z=1.39$. The C-M relation (the solid red line in
  Fig.~\ref{fig:HAWKI_CMD}) is determined from galaxies that lie
  within this circle.}

\label{fig:HAWKI}
\end{figure*}

\subsection{Object Detection and Photometry}

The analysis is split into two regions: the central arc-minute of the
cluster, which is common to all frames, and the region that is outside
of this.

For the central arc minute of the mosaic, we used version 2.5.0 of
SExtractor \citep{Bertin96} in double image mode to detect objects in
the unsmoothed Ks-band image and to measure aperture magnitudes in the
smoothed and aligned images. The aperture diameter is set to 9 pixels
(0\farcs96 or 8\,kpc at the redshift of the cluster), which is about
twice the FWHM of point sources in the J-band image. We use the 
SExtractor neural network classifier to separate stars from galaxies.

For regions outside the central arc minute we used SExtractor in
single image mode to measure aperture magnitudes in the smoothed and
aligned images. To simplify the analysis and to reduce errors caused
by slight misalignments, the aperture is set to 12 pixels.

For both the central region and the region outside of it, colours are
estimated from aperture magnitudes.

Total magnitudes are estimated differently. Given the crowded nature
of the centre of the cluster, we use version 2.0.3c of GALFIT
\citep{Peng02} on the unsmoothed Ks-band images to estimate total Ks
magnitudes. In crowded regions, the MAG\_AUTO estimate of SExtractor
systematically overestimates the flux of galaxies that lie within the
detection isophote of a brighter neighbour. For isolated galaxies,
GALFIT and SExtractor MAG\_AUTO magnitudes agree with a dispersion of
0.2 magnitudes.

The central wavelength of the HAWK-I J filter is slightly redder than
the 2MASS J filter, so we expect a small colour term when transferring
from HAWK-I colours to 2MASS colours. To estimate the colour term, we
used the transmission curves of the
HAWK-I\footnote{http://www.eso.org/instruments/hawki/} and
2MASS \citep{Cohen03} filters and a range of
theoretical galaxy spectral energy distributions \citep{BC03}. We find

\begin{equation}
(J-Ks)_\mathrm{2MASS} = 1.032\,(J-Ks)_\mathrm{HAWK-I} - 0.025
\label{eq:1}
\end{equation}

Note that this transformation is strictly valid for the spectral
energy distributions (SED) that were used to compute it. It is
unlikely to be valid for other SEDs.  In particular, late type stars,
because of their broad spectral features, will require a different
relation.

The Galactic extinction along the line of sight to
\object{XMMU\,J2235.3-2557} was estimated from the dust maps of
\citet{Schlegel98}. The extinction is 0.019 and 0.008 magnitudes
in J and Ks, respectively, and both colours and total magnitudes were
corrected for it.

\subsection{Photometric errors}

In order to obtain a reliable estimate of the location of the red
sequence and the scatter within it, it is important to quantify random
and systematic errors.

In the fully processed images, the noise between pixels is correlated.
This is caused by the interpolation that is used when aligning images,
and by large scale features in the sky background \citep{Labbe03}. The
end result is that the dispersion in the flux integrated over a fixed
aperture is underestimated if one simply scales the noise in a single
pixel by the square root of the number of pixels in the aperture.

To overcome this difficulty, the error in the flux in an aperture
of a certain size is estimated by randomly placing these apertures in
regions that are free of detectable objects and by computing the
dispersion in the integrated flux. The dispersion is then combined
with the object flux to compute a magnitude error.

In addition to random errors caused by photon shot noise, there are
systematic errors that can affect the location of the red sequence in
the C-M diagram and the scatter within the sequence itself. We discuss
likely sources of systematic error and assess their impact.

The correction that is applied for the difference between the HAWK-I
and 2MASS filters (Eq. \ref{eq:1}) is possibly one of the largest
sources of systematic error. However, without observational data to
validate the correction that is applied in Eq.~\ref{eq:1}, it is
difficult to assess the accuracy of the correction and to assign a
systematic error. So, we do not assign an error, but note that it
might be significant if the filters are very different to those that
are published, or if the spectra of high redshift passively
ellipticals are significantly different to the models used to derive
Eq.~\ref{eq:1}.

An error in one or both zero points can shift the entire C-M diagram
either up or down. By using 2MASS stars in the field (see
Sec.~\ref{sec:data}), we measured a mean colour difference of 0.009
magnitudes between the colour corrected HAWK-I colours and the 2MASS
colours. We do not correct for this offset.

Another source of systematic error comes from the matching of the PSFs
in the J and Ks-band images. An error in matching the PSFs will result
in a offset in the colours, since the colours are computed from fixed
apertures. We evaluate the likely size of this offset by comparing the
colours of bright unsaturated stars in two different apertures: a 12
pixel diameter aperture and a larger 4\arcsec radius aperture. The average
offset is 0.015 magnitudes with no evidence for a dependence on where
the star was located in field-of-view. When this offset was larger
than 0.02 magnitudes (just one case), we corrected the colours by this
offset. Hence, a suitable value for this error is 0.02 magnitudes.

Although the resolution of these images are excellent for images that
are taken from the ground, faint galaxies that lie near to the line of
sight to brighter galaxies might go undetected. It is difficult to
quantify the effect of these galaxies, but, qualitatively, since they
are likely to be bluer, they will bias the colours of the red sequence
galaxies to bluer colours.

\section{The C-M diagram}

The region surrounding the centre of \object{XMMU\,J2235.3-2557} is shown
in Fig.~\ref{fig:HAWKI}. The centre of the cluster and, in particular, the
brightest cluster galaxy (BCG) are very conspicuous in the HAWK-I
near-IR images. About half of the IR light in this central region
comes from the BCG.

\begin{figure*}
\centering
\includegraphics[width=17cm]{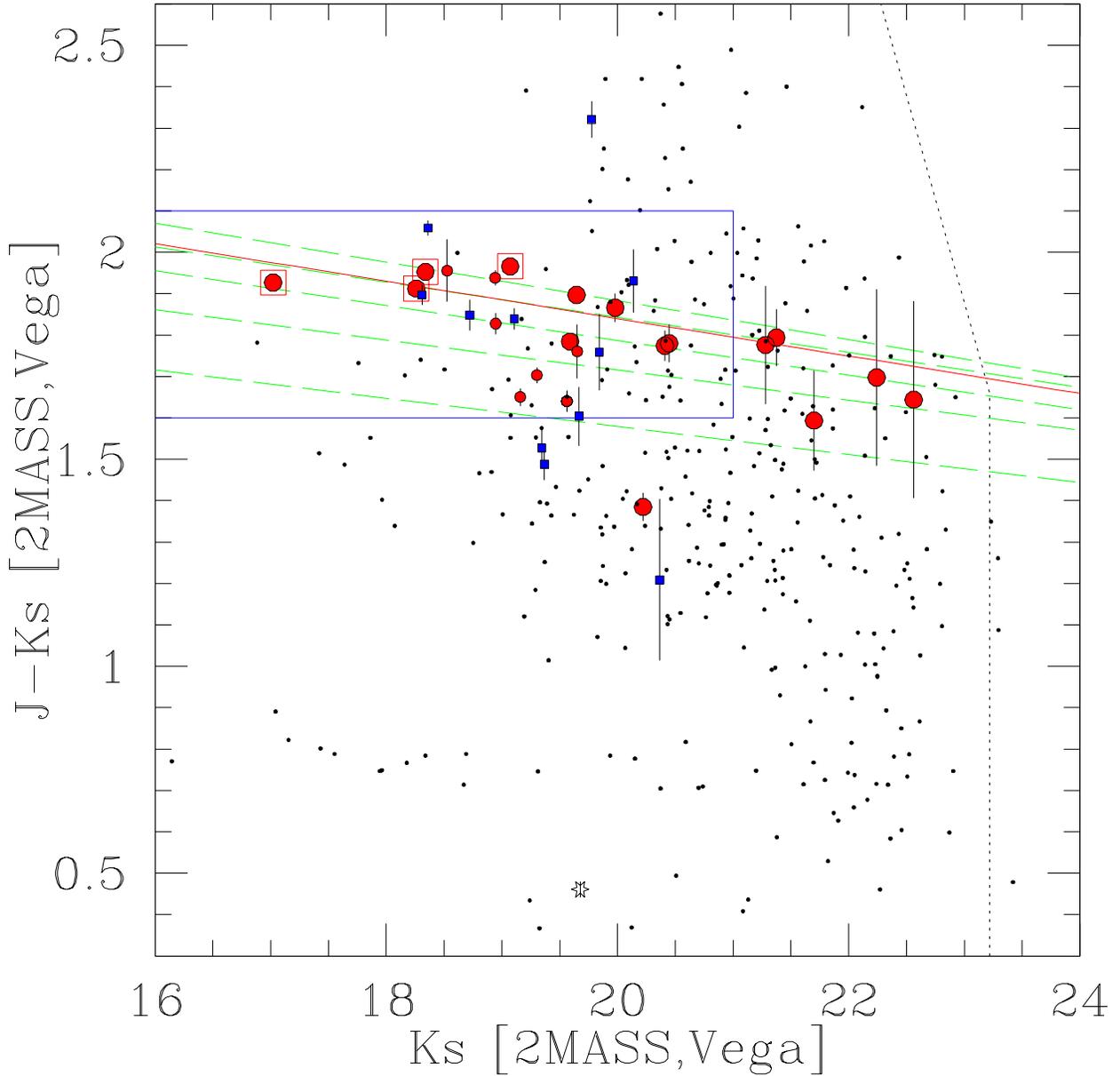}
\caption{Colour magnitude diagram of
  \object{XMMU\,J2235.3-2557}. Galaxies in the core of the cluster
  (within the dotted circle that is shown in Fig.~\ref{fig:HAWKI}) are
  plotted as large red circles. The continuous red line is a fit to
  the large red symbols that lie within the blue box. Four of these
  galaxies are spectroscopically confirmed cluster members and are
  replotted as open squares. The dotted lines represent the detection
  limits reported in Table \ref{table:observations}. The
  near-horizontal dashed green lines are the result of evolving the
  red sequence of the Coma cluster backwards in time for different
  formation redshifts. From top to bottom, the formation redshifts are
  5, 4, 3, 2.5 and 2. Cluster members that are outside the core are
  plotted as small red circles (for those that have no detectable
  [OII] emission) or small blue squares (for those that have [OII]
  emission). See the main text for details.}

\label{fig:HAWKI_CMD}
\end{figure*}

The colour magnitude diagram of selected targets in a region centred
on the cluster is shown in Fig.~\ref{fig:HAWKI_CMD}. Galaxies within
100 pixels (90\,kpc) of the centre of the cluster (within the dotted
circle that is drawn in Fig.\ref{fig:HAWKI}) are plotted as large red
circles. Of these, four are spectroscopically confirmed cluster
members\footnote{In this paper, we define galaxies with redshifts that
lie within the redshift interval $1.375 < z < 1.400$ as cluster
members.} and are replotted with open red squares. The
[OII]\,$\lambda\lambda$\,3727 emission line is not detected in any
these galaxies. Within this central region, there is one star, which
is plotted as such in Fig. \ref{fig:HAWKI_CMD}.

The continuous red line is a fit to the galaxies that lie within the
central region and within the blue rectangle. The fit and the
intrinsic scatter were determined by adjusting the intrinsic scatter
until the reduced $\chi^2$ of the fit was one. Since the errors in the
photometry are small, the intrinsic scatter is very similar to the
measured one and is $0.055 \pm 0.018$ magnitudes. A Monte-Carlo
simulation was used to compute the error in our estimate of the
intrinsic scatter.  We ran 1000 realisations of the data with
properties identical to the real one (magnitudes and photometric
errors) and treated these realisations in the same way as the real
data. The error was then estimated from the distribution of the
fitted intrinsic scatter.

Our estimate of the intrinsic scatter is not very sensitive to the
exact size of the region that is used to include galaxies.  Doubling
or halving the size of the region changes the intrinsic scatter by
less than 0.01 magnitudes. Nor is it very sensitive to the precision
at which we have estimated photometric errors. Doubling the size of
the photometric errors reduces the intrinsic scatter by 0.01
magnitudes.

Although not all of the galaxies within the blue rectangle are
spectroscopically confirmed cluster members, it is likely that most of
them are cluster members. Using galaxies within the 13\farcm5 x
13\farcm5 field of view that was imaged by HAWK-I, we estimate that
one galaxy would, by chance, lie within 100 pixels of the
cluster centre and within the limits specified by the blue rectangle.

Outside this central region, spectroscopically confirmed cluster
members with and without [OII] emission (11 and 7 members,
respectively) are plotted as the small blue squares and small red
circles, respectively. Isolated objects (objects that SExtractor does
not identify as a merged detection) within 0.03 degrees (0.9\,Mpc) of
the cluster centre and generally without spectroscopic redshifts are
plotted as black dots. The size of this region was chosen as a balance
between having enough points to illustrate the colour of objects
outside the cluster and obscuring the plot with too many points.  The
objects that make up the slightly tilted sequence at $J-Ks \sim 0.8$
are most likely evolved late type stars in the Galaxy. If one were to
increase the size of the region, a second, less well defined sequence
appears at $J-Ks \sim 0.4$.

%
%______________________________________________________________

\section{Discussion}

Galaxies within 90\,kpc of the centre of the
\object{XMMU\,J2235.3-2557} lie on a well defined red sequence. The
slope ($-0.045 \pm 0.019$) and the scatter ($0.055 \pm 0.018$ magnitudes) are
similar to those measured in the X-ray luminous galaxy cluster
\object{RDCS~J1252.9-2927} at z=1.24 \citep{Lidman04, Demarco07}.

Outside this central region the red sequence is far less well
defined. Although the number of spectroscopically confirmed cluster
members is rather modest and one must bear in mind the selection bias
that comes from the relative ease in measuring the redshifts of star
forming galaxies, it is clear that the galaxies outside the cluster
centre are considerably more active. The [OII] emission line is seen
in more than half of these galaxies.

Cluster members with [OII] emission have a broad colour distribution,
from $J-Ks \sim 1.2$ to $J-Ks \sim 2.3$. Some of the [OII] cluster
members land on the red sequence, thus agreeing with the notion that
the [OII] emission line is a more sensitive indicator of active star
formation or AGN activity \citep{Yan06} than galaxy colour. The two
reddest cluster members are [OII] emitters and both are significantly redder
than the red sequence, perhaps indicating the presence of significant
amounts of dust in these galaxies.  Outside the cluster centre, the
two brightest cluster members are also [OII] emitters.

On the other hand, cluster members that are outside the cluster centre and
without detectable [OII] emission either land on the red sequence or
slightly blue of it, perhaps indicating that some of these galaxies are
significantly younger than galaxies in the central part of the
cluster.  Alternatively, these galaxies could be bluer because there
was a burst of star formation that ended just before these galaxies
were observed. Photometry in bluer pass-bands and/or deep spectroscopy
of these galaxies can be used to distinguish between these two
possibilities.

Mullis et al. (2005) noted that two galaxies (one in the cluster core
and another in the cluster outskirts) might be located in the
foreground of the cluster.  Removing these galaxies from the sample,
does not significantly change the results.

\subsection{The location of the red sequence}

The location of the red sequence was modelled with the Bruzual \&
Charlot (2003) population synthesis code.  We adopted a simple stellar
population (SSP) model with a Salpeter initial mass function, Padova
1994 evolutionary tracks and no dust extinction.  We did not model
more complex star formation histories because of the limited colour
information available (for a more extended analysis, see Rosati et
al.~in preparation). In order to reproduce the slope of the red
sequence, the red sequence in the Coma cluster \citep{Bower92} was
fitted with SSP models of varying metallicities.  The model red
sequence was then passively evolved back in time to $z=1.39$. Model
red sequences were computed for 5 different formation redshifts ($z_{\rm f}$=5,
4, 3, 2.5 and 2). They are plotted as the dashed green lines in
Fig. \ref{fig:HAWKI_CMD}. Note that the modelling is based on the
assumption that the slope of the red sequence is entirely due to the
mass-metallicity relation.  The close match between the observed slope
(the continuous red line in Fig.\ref{fig:HAWKI_CMD}) and computed ones
(the dashed green lines in Fig.\ref{fig:HAWKI_CMD}) suggests that this
is a reasonable assumption.

Galaxies within the central 180\,kpc are clearly old. The models
suggest that the average formation redshift of the stars in these
galaxies is $z_{\rm f} \sim 4$. As discussed in the previous section,
there is a systematic uncertainty of a few hundredths of a magnitude
in the colours, which could move the average formation redshift to as
early as $z_{\rm f}=4.5$ or as late as $z_{\rm f}=3.5$. This redshift
is similar to the estimated formation redshifts of distant clusters
that are at slightly lower redshifts. In \object{RDCS~J1252.9-2927} at
$z=1.24$, for example, it is estimated that the average formation
redshift is $z_{\rm f} \sim 3.5$ -- $4$ \citep{Blakeslee03,Gobat08}.

Outside the central region, cluster members appear to be, on average,
younger. More than half these galaxies show signs of active star
formation and about half of those that do not are significantly bluer
than galaxies on the red sequence. If we were to interpret the
difference in colour as a difference in age, then these galaxies are
about 1.5\,Gyr younger.

One can compute the redshift at which the colours of these galaxies
would become indistinguishable from the colours of galaxies on the
red sequence if all the galaxies were allowed to evolve passively. Using the
models that were used to fit the location of the red sequence, this occurs
at $z \sim 1$, i.e., 1.3\,Gyr later. 

Alternatively, if the blue colours are caused by the presence of
a young stellar population from a recently truncated episode of star
formation, then these galaxies will join the red sequence more quickly.

The contrast between the cluster core, which is dominated by passively
evolving galaxies on the red sequence, and the cluster outskirts, which is
dominated by active galaxies spanning a wide range of colours,
suggests that the red sequence of this cluster is being built from the
dense core of the cluster to relatively sparse outskirts.  Furthermore,
the process is a quick one. Many of the galaxies not currently
on the red sequence will be on the red sequence by $z=1$.

\subsubsection{A flattening of the C-M relation at the bright end?}

The colours of the four brightest galaxies in the centre of the
cluster, which cover a range in brightness of a factor of five, are
very similar, suggesting that the slope of the C-M relation is flat at
the bright end. A flattening in the C-M relation had also been noted
in the X-ray luminous cluster \object{RDCS~J1252.9-2927} at z=1.24
\citep{Demarco07}. If the flattening is real, then the C-M relation at
fainter magnitudes becomes steeper.

At this stage, the flattening is only suggestive, since the number of
bright galaxies over which it can be measured is small. Additional
observations of other high redshift clusters will be required to confirm
or refute the trend.

Dry red mergers - the merging of two passively evolving galaxies
without subsequent star formation - is a process that can lead to a
flattening of the C-M relation. For example, a 1:1 merger will result
in a galaxy that is 0.75 magnitudes brighter and displaced by 0.035
magnitudes blue-wards of the C-M relation.

However, if dry red mergers were the cause for the flattening at the
bright end, then one would expect this structure to persist as the
cluster evolves, as all the galaxies at the bright end would have
similar metallicities and ages. Although a flattening at the bight end
of the C-M relation is evident in some semi-analytic N-body
hierarchical galaxy formation models \citep{deLucia04a,Menci08}, it
has not been observed in well studied clusters at lower redshifts
\citep[e.g.,][]{Terlevich01,Andreon06,Eisenhardt07}.

Alternatively, wet mergers - the merging of galaxies with subsequent
star formation - may be responsible for part of the apparent
flattening. In Fig.~\ref{fig:HAWKI_CMD}, the BCG is slightly bluer
than expected. Since the [OII] line was not observed in the spectrum
of the BCG, the slightly blue colours could be the result of recent
but now extinguished burst of star formation that was triggered by a
merger. Over time, the effect of this burst on the colour will fade and the
colour of the BCG will move back towards the C-M relation. 

\subsubsection{Truncation of the red sequence}

The continuous red line in Fig.~\ref{fig:HAWKI_CMD} is a fit to the
colour of galaxies in the core of the cluster.  Although only galaxies
within the blue box of Fig.~\ref{fig:HAWKI_CMD} were used in the fit,
there are galaxies outside this box that also lie close to this line.

Bearing in mind that some fainter galaxies might go undetected because
of the crowded nature of the cluster centre, that not all of the
galaxies plotted as large symbols in Fig.~\ref{fig:HAWKI_CMD} have
been confirmed as cluster members, and that the number of galaxies in
this plot is quite small, there is no evidence that the red sequence
is strongly truncated at $\mathrm{Ks}\sim 20.5$, as was detected in
the large scale structure surrounding \object{RDCS~J1252.9-2927} at
z=1.24 \citep{Tanaka07} and in several proto-clusters
\citep{Kodama07}.

Hence, in this property, the core of \object{XMMU\,J2235.3-2557} is
similar to the core of \object{RDCS~J1252.9-2927}. However, it remains to
be tested if the luminosity function of galaxies within the large scale
structure surrounding the core of \object{XMMU\,J2235.3-2557} is
truncated, as has been observed in \object{RDCS~J1252.9-2927}.

\subsection{The intrinsic scatter}

While the tilt of the red sequence is largely due to a relationship
between metallicity and luminosity and hence mass (brighter and larger
galaxies are more metal rich) \citep{Kodama99}, the reason for the
finite intrinsic scatter is less clear. It could be caused by
variations in age, metallicity or a combination of both. Dust could
also play a role. For early-type galaxies in the field, the scatter is
caused by differences in age and metallicity \citep{Gallazzi06}, with
variations in metallicity playing an increasingly important role as
the mass of the galaxy increases.

The intrinsic scatter in the observer frame $\mathrm{J-Ks}$ colour of
cluster galaxies about the C-M relation of
\object{XMMU~J2235.3-2557} is $0.055 \pm 0.018$\,magnitudes. Since
only a few of the galaxies that were used to measure the scatter have
been spectroscopically confirmed, it is possible, although unlikely,
that the estimate of the scatter has been inflated by non-cluster
members.

This scatter is similar to the scatter in observer frame
$\mathrm{J-Ks}$ colours for galaxies in the X-ray luminous cluster
\object{RDCS~J1252.9-2927} at z=1.24. \citep{Lidman04,Demarco07}. It
is also similar to the scatter in the observer frame $\mathrm{J-K}$
colour measured for galaxies in clusters at lower redshifts
\citep{Stanford98,Holden04}. Note, however, that the scatter reported
here is in the observer frame, so the scatters are not directly
comparable.

At $z=1.39$, observer frame J and Ks approximately correspond to rest
frame V and z. For Coma, the measured scatter in the V-I colour, the
nearest colour for which the intrinsic scatter has been estimated, is
0.031 magnitudes~\citep{Eisenhardt07}. In detail, converting the intrinsic
scatter from observer frame $\mathrm{J-Ks}$ to rest frame
$\mathrm{V-I}$ or visa-versa requires an assumption of what causes
the scatter (e.g., age, metallicity, or a combination of both) and a
model of how this affects the spectral energy distribution.  If the
cause of the scatter is due to metallicity, for example, then, by
using SSP models from \citet{BC03}, the inferred scatter in rest frame
$\mathrm{V-I}$ for galaxies in the core of \object{XMMU~J2235.3-2557}
can be estimated and is found to be $0.035\pm0.011$\,magnitudes, which is
the same as that measured in Coma.

As seen in other studies \citep[][Mei el al. in
preparation]{Stanford98,Blakeslee03}, the size of the scatter is
remarkably constant. In this study, the scatter is constant over a
time interval spanning 8.8\,Gyr, which is a significant fraction of
the age of the Universe. However, we should be aware of possible
biases that are caused by the way low and high redshift samples are
selected. In this particular case, the scatter measured for Coma was
measured over a region that was several times larger than the region
used for \object{XMMU~J2235.3-2557} ($\sim1$\,Mpc versus $\sim
0.2$\,Mpc).  Even if similar size regions had been selected, it is
likely that some galaxies that were outside the central region at
$z=1.39$ would by now lie within the central region and possibly
visa-versa. This is a form of progenitor bias \citep{vanDokkum01},
where the progenitors of youngest present day early type galaxies drop
out of high redshift samples.

Of particular interest are cluster members that are significantly
redder than the C-M relation. The most extreme example without [OII]
in emission in \object{XMMU~J2235.3-2557} has a colour that is 0.08
magnitudes redder.  The colour offset can easily be accounted for by
an increase in the metallicity. In this particular example, an
increase in the metallicity from $Z=0.017$ to $Z=0.023$ would account
for the redder colour. On the other hand, if the redder colour was due to
the galaxy being older and if errors have not been underestimated,
then this galaxy would have to be as old as the Universe at $z=1.39$.

\subsection{Comparison with hierarchical galaxy formation models}

The approach to modelling the behaviour of baryons in N-body
hierarchical galaxy formation models comes in two main flavours:
semi-analytical approaches \citep[e.g.][]{Kauffmann98,deLucia04a,Menci08}, and
hydrodynamical approaches \citep[e.g.][]{Romeo08}.

Here we make a qualitative comparison between the results from these
models and the observations. A quantitative comparison requires
detailed understanding of the quantities that are being measured in
both the observations and the simulations. Such a comparison is beyond
the scope of this paper.
 
In the hydrodynamical simulation run by Romeo et al. (2008), the
fraction of galaxies actively forming stars strongly depends on
environment and galaxy mass. For galaxies more massive than
$2\times10^{10}\,M_{\sun}$, star formation first stops in the centre of the
most massive clusters and then stops in the cluster outskirts or the
centres of groups later.

In \object{XMMU\,J2235.3-2557}, which has an X-ray temperature that is
slightly higher than C2 model of \citet{Romeo08} (8-9\,kev versus
6\,kev), none of the galaxies in the core of the cluster are actively
forming stars, whereas about half of the galaxies outside the core
are. This picture is broadly consistent with the N-body
simulations. However, we caution that the census of star forming
galaxies in the centre and the outskirts of the cluster is far from
complete.

There is, however, a notable difference between the models in Romeo
et al. and the observations of \object{XMMU\,J2235.3-2557}. The slope
of the red sequence in the models flattens with increasing cluster
redshift and actually becomes positive. The flattening with redshift
had also been noted in semi-analytical hierarchical simulations
\citep{Kauffmann98,Menci08}. For galaxies in the core of
\object{XMMU\,J2235.3-2557}, the slope of the red sequence is not
positive, nor even flat. It is clearly negative.

As shown above, a simple model in which the galaxies form at very high
redshift and evolve passively thereafter fits the slope of the
C-M relation very well.

The semi-analytical and hydrodynamical simulations are also unable to
reproduce both the size of the intrinsic scatter and the lack of
evolution in the scatter with redshift.

In early semi-analytical simulations \citep{Kauffmann98}, the scatter
decreases with increasing redshift because the selection of galaxies
in high redshift clusters is biased towards galaxies that have already
formed at very high redshifts. From redshift zero to redshift 1.5, the
scatter in the rest frame U-V decreases by a factor of $\sim 2$
\citep{Kauffmann98}. In hydrodynamical simulations \citep{Romeo08}, a
similar decrease in the scatter is found for galaxies that lie on the
dead sequence -- galaxies that are no longer forming stars.

In more recent semi-analytic simulations \citep{Menci08}, the average
scatter up to $z=1.5$ is almost independent of redshift, but is a
factor of two to three larger than the observed value. Part of the
discrepancy might come from the way red sequence galaxies are selected
\citep{Menci08}. In \object{XMMU\,J2235.3-2557}, however, all but one
of the galaxies within the core of the cluster were used to compute
the scatter, so the scatter in the core of this cluster is clearly
small and consistent with Coma.

The scatter in the models of \citet{Menci08} have a large range and
the observations tend to land at the lower boundary of that
range. This may indicate that the simulations consider a broader range
of clusters and that the observations are biased to the most evolved
examples. Alternatively, it might mean that the models are missing
physical processes that lead to the small scatter.

\section{Summary and Conclusions}

We have presented near-IR observations of the X-ray luminous cluster
{XMMU\,J2235.3-2557} at $z=1.39$ and we have built a C-M diagram of
objects inside and outside of the core of the cluster. While galaxies inside
the cluster core form a well defined red sequence with no evidence of
ongoing star formation, cluster members outside the core are much more
diverse.

The colour of galaxies inside the core can be matched with SSP models
that are $\sim 3$ Gyr old, corresponding to a redshift of formation
of $z_{\rm f}\sim 4$. These galaxies are already very old, especially
when we consider that the cluster was observed at a time when the age of
Universe was 4.6\,Gyr.

Cluster members outside the core do not form a well defined red sequence.
Over half these galaxies are forming stars and some of these are
either considerably redder than the red sequence, perhaps 
indicating the presence of dust, or considerably bluer. The other
half do not appear to be forming stars, but are, on average, displaced
towards bluer colours, perhaps indicating that they either stopped
forming stars recently or are younger than galaxies on the red sequence.

The contrast between the cluster core, which consists of a population
of evolved galaxies with uniform colours, and the cluster outskirts,
which consists of a population of active galaxies with diverse
colours, suggests that the red sequence of this cluster is being built
from the inside to the outside or, alternatively, from the dense core to
the relatively sparse outskirts.

\begin{acknowledgements}

We acknowledge the very useful correspondence that we had with Andreas
Seifahrt concerning the processing of HAWK-I data. This publication
makes use of data products from the Two Micron All Sky Survey, which
is a joint project of the University of Massachusetts and the Infrared
Processing and Analysis Center/California Institute of Technology,
funded by the National Aeronautics and Space Administration and the
National Science Foundation. This research was supported by the
Deutsche Forschungs Gesellschaft cluster of excellence program Origin and
Structure of the Universe (www.universe-cluster.de).

\end{acknowledgements}

\bibliography{0528}

\end{document}